\documentclass[%
reprint,
 showpacs,
amssymb,
aps,
pre,
 longbibliography
]{revtex4-1}

\usepackage{booktabs} 

\usepackage{amsmath,dsfont}
\usepackage{mathbbol}
\usepackage{bbold}
\usepackage{mathrsfs}
\usepackage{graphicx}
\usepackage{dcolumn}
\usepackage{bm}
\usepackage[usenames]{color}
\usepackage[normalem]{ulem}
\usepackage{hyperref}
\usepackage{comment}
\usepackage{amsthm}
\usepackage{listings}

\usepackage[per-mode=symbol]{siunitx}   
\usepackage[version=3]{mhchem}

\newcommand{\ie}{i.\,e.}%

\newcommand{\ket}[1]{\left|#1\right\rangle}

\newcommand{\ucm}{\affiliation{Departamento de Qu\'imica F\'isica, Universidad Complutense de Madrid, 28040 Madrid, Spain}}%

\begin{document}

\title{Quantum Optimal Control at Intermediate Times: Controlling Revivals in Spin Chains}

\author{Juan J.\ Omiste}\email{jomiste@ucm.es}\ucm
\author{Ignacio R. Sol\'a}\ucm

\date{\today}
\begin{abstract}

Standard Quantum Optimal Control (QOC) protocols typically maximize a physical objective just at a final target time. However, tracking or measuring the quantum state during the time evolution requires the control at intermediate stages of their evolution. In this work, we extend QOC to accommodate the simultaneous optimization of observables at arbitrary intermediate times. Using a variational approach, we show that intermediate observations induce discontinuities in the costate trajectory, which can be handled with a Krotov algorithm leading to monotonic optimization and convergent results in the limit of zero temporal measurement windows. We apply this multi-time formulation to a Heisenberg XXX spin chain to control the propagation, including field-free revivals, of a Dicke state excitation. Our results demonstrate that simultaneous optimization of the same observable reshapes the driving field to balance intermediate targets with final populations. Finally, we show how this framework enables dynamic tracking of spin excitations and the active manipulation of post-pulse quantum state revivals.

\end{abstract}

\maketitle

\section{Introduction}
\label{sec:introduction}

The precise manipulation of quantum systems lies at the heart of modern quantum information processing, quantum simulation, and quantum state engineering~\cite{Koch2022}. High-fidelity control is one of the main goals of Quantum Optimal Control (QOC) theory, providing the mathematical framework necessary to design driving fields that guide a system toward a desired objective with maximum efficiency~\cite{Peirce1988,Kosloff1989}. Many areas have benefit from it, with recent applications in chirality ~\cite{Berggotz2025,Leibscher2022a}, 
ultracold matter~\cite{Sklarz2002,Bohn2017,Li2023,Smucker2024,Inarrea2024,Hutson2024,Kaufman2024UltracoldMolecules}, 
spin dynamics~\cite{Khaneja2001PRA,Khaneja2001CP,Khaneja2005,Skinner2005}
or Rydberg atoms~\cite{Goerz2014,Browaeys2020,Bluvstein2021,Jandura2022,Sola_Nanoscale2023,Sola_PRA2023,Sola_AIPadv2023,Carrera2025,Carrasco2512.07656}.
Among the various numerical approaches developed to solve the underlying Euler-Lagrange equations~\cite{Castro2025}, Krotov's method and their variants, are widely used for their monotonic convergence~\cite{Krotov1983,Somloi1993,Zhu1998a,Zhu1998b,Palao2002,Palao2003,Maday2003,Ohtsuki2004,Schirmer2007,Ohtsuki2007,Ohtsuki2008,Ohtsuki2021,Ho2010,Liao2011,Goerz2019a}
and flexibility in handling diverse physical constraints~\cite{Riviello2015,Shu2016a,Shu2016b,Goerz2019a,Werschnik2007}.

Conventionally, standard QOC protocols are formulated to optimize the expectation value of an operator at a single measurement time~\cite{Peirce1988,Werschnik2007}. However, emerging quantum technologies increasingly demand a more complex control landscape in which the system must be monitored, protected, or actively manipulated in intermediate stages of its evolution~\cite{Magann2023}. Examples include multi-step quantum logic gates, the prevention of leakage into unwanted subspaces during a pulse, and the precise timing of wavepacket revivals. While continuous time-dependent tracking can be implemented, many realistic protocols require maximizing discrete, instantaneous observations at arbitrary intermediate times. 

Recently, the canonical solutions of quantum optimal control theory were revisited using a variational approach~\cite{Castro2025}, highlighting the discontinuous nature of the costate at the measurement time, in contrast to the imposition of continuity~\cite{Werschnik2007,Peirce1988}. Here we extend that framework to the case of multiple observables at arbitrary intermediate times, treating them as Dirac-delta window functions. These conditions imply a jump in  the costate due to the inhomogeneity appearing in the Schr\"odinger-like equation driving the costate. Furthermore, we provide a straightforward implementation within Krotov's method. Furthermore, we rigorously validate the numerical convergence of this approach by proving that it represents the smooth mathematical limit of sequentially narrowing, finite-width temporal measurement windows.

To benchmark the utility and explore the physical implications of this multi-time control protocol we explore the tracking of transport within a five-spin-1/2 chain modeled by the Heisenberg XXX model restricted to the single-excitation manifold. Specifically, we investigate the tracking of observables during the dynamics including postpulse control of field-free population dynamics of a target Dicke state. 

The structure of this paper is organized as follows. In Sec.~\ref{sec:system_and_methodology}, we introduce the general variational formalism, derive the costate jump conditions, and outline their implementation and convergence within Krotov's algorithm, alongside a description of the spin-chain model. In Sec.~\ref{sec:results}, we present and discuss the numerical results for both finite-width windows and post-pulse field-free revival control. Finally, our conclusions and future outlook are summarized in Sec.~\ref{sec:conclusions_and_outlook}.

\section{System and Methodology}
\label{sec:system_and_methodology}
In this section, we present an extension of the Quantum Optimal Control (QOC) framework to maximize the expectation value associated to several observables at different times, represented by $O(t)$. 

\subsection{General formalism}
We set the quantum system in the time interval $[0,T_\text{end}]$ and aim to maximize the time dependent observable $O(t)$. Following the Quantum Optimal Control variational method~\cite{Castro2025}, 
we aim to maximize the functional given by~\cite{Kosloff1989,Werschnik2007}
\begin{eqnarray}
    \nonumber
J[\psi,\psi^*,\chi,\chi^*,\varepsilon]&=&J_\text{opt}[\psi,\psi^*]+J_\text{cost}[\varepsilon]\\
\label{eq:jtotal}
&&+ J_\text{TDSE}[\psi,\psi^*,\chi,\chi^*,\varepsilon]
\end{eqnarray}
where 
\begin{equation}
    \label{eq:optimize}
    J_\text{opt}[\psi,\psi^*]=\int_0^{T_\text{end}} \int_\Omega \psi^*(\Omega, t)O(t)\psi(\Omega, t)\mathrm{d}\Omega\mathrm{d}t,
\end{equation}
is associated with the observable $O(t)$, and
\begin{equation}
    \label{eq:cost}
    J_\text{cost}[\varepsilon]=-\int_0^{T_\text{end}}\cfrac{\alpha}{S(t)} \varepsilon(t)^2\mathrm{d}t,
\end{equation}
accounts for the fluence of the driving pulse~$\varepsilon(t)$ and is constrained by a penalty factor $\alpha$ and shaped with the envelope $S(t)$. Finally, adding the Lagrange multiplier $\chi(\Omega,t)$ in the term
\begin{eqnarray}
    \nonumber
  &&  J_\text{TDSE}[\psi,\psi^*,\chi,\chi^*,\varepsilon]=\\
  \nonumber
&&-2\operatorname{Im}\int_0^{T_\text{end}}\int_\Omega\chi^*(\Omega,t)\left(i\frac{\partial}{\partial t}-H(\Omega, \varepsilon)\right)\psi(\Omega,t)\mathrm{d}\Omega\mathrm{d}t \ ,
\end{eqnarray}
the Time-Dependent Schr\"odinger equation (TDSE) is fullfilled by $\psi(\Omega,t)$. 

The extremal solutions which maximizes the functional 
$J[\psi,\psi^*,\chi,\chi^*,\varepsilon]$ for fixed values of $\psi(\Omega,t)$ and $\chi(\Omega,t)$ at $t=0$ and $T_\text{end}$, fulfills the Euler-Lagrange equations~\cite{Castro2025}
\begin{subequations}
    \begin{equation}
        \label{eq:tdse_psi}
        i\dfrac{\partial}{\partial t}\psi(\Omega,t)-H(\Omega,\varepsilon)\psi(\Omega,t)=0,
    \end{equation}
    \begin{equation}
        \label{eq:tdse_chi}
        i\dfrac{\partial}{\partial t}\chi(\Omega,t)-H(\Omega,\varepsilon)\chi(\Omega,t)=-iO(t)\psi(\Omega,t),
    \end{equation}
    \begin{equation}
        \label{eq:field_ele}
        \varepsilon(t)=\dfrac{S(t)}{\alpha}\operatorname{Im}\int_\Omega\chi^*(\Omega,t)\dfrac{\partial}{\partial \varepsilon}H(\Omega,\varepsilon) \psi(\Omega,t),
    \end{equation}
    \begin{equation}
    \label{eq:psi_boundary}
        \psi(\Omega,t=0)=\psi_0,
    \end{equation}
\end{subequations}
and their complex conjugates, and the transversality condition
    \begin{equation}
        \label{eq:chi_boundary}
        \chi(\Omega,T_\text{end})=0
    \end{equation}
Therefore, the extremal solution fulfills the set of equations~\eqref{eq:tdse_psi}-\eqref{eq:psi_boundary} and~\eqref{eq:chi_boundary}~\cite{Werschnik2007}.
\subsection{Extending the control to intermediate times}
In order to maximize $N_\text{obs}$ observables at different intermediate times $T_k\in(0,T_\text{end})$, we take the observable $O(t)$ as 
\begin{equation}
    \label{eq:obs_k}
O(t)=\sum\limits_{k=0}^{N_\text{obs}}O_k\delta(t-T_k),
\end{equation}
where $\delta(t)$ stands for the Dirac delta and $O_k$ is a time-independent observable. In this case, Eqs.~\eqref{eq:tdse_psi} and \eqref{eq:tdse_chi} become the usual TDSE for $\psi(\Omega,t)$ and $\chi(\Omega,t)$ except at the intermediate times $T_k$, where the discontinuities in the costate have to be dealt with appropriately. To handle this problem, we refer to the procedure in Ref.~\cite{Castro2025}, where we compute the integral at time around $t=T_k$. We find, for $\epsilon>0$
\begin{equation}
    \label{eq:tk_discontinuity}
    \lim\limits_{\epsilon\rightarrow 0}\chi(\Omega,T_k-\epsilon)=\lim\limits_{\epsilon\rightarrow 0}\chi(\Omega,T_k+\epsilon)+O_k\psi(\Omega,T_k).
\end{equation}
 In the following, we discuss the implications of this condition on the QOC equations.

\subsubsection{Single measurement time}
Let us consider first the single measurement case, $O(t)=O\delta(t-T)$. Inserting this observable in Eq.~\eqref{eq:tk_discontinuity}, we obtain
\begin{equation}
    \label{eq:single_observable}
    \lim\limits_{\epsilon\rightarrow 0^+}\chi(\Omega,T-\epsilon)=\lim\limits_{\epsilon\rightarrow 0^+}\chi(\Omega,T+\epsilon)+O\psi(\Omega,T).
\end{equation}
Since $\chi(\Omega,t)$ fulfills the TDSE in the interval $t\in (T,T_\text{end})$, $\chi(\Omega,T+\epsilon)=0$ for $\epsilon>0$, Therefore, we obtain, for $\chi(\Omega,T-\epsilon)$, 
\begin{equation}
    \label{eq:single_observable_chi}
    \lim\limits_{\epsilon\rightarrow 0^+}\chi(\Omega,T-\epsilon)=O\psi(\Omega,T),
\end{equation}
which is equivalent to the standard boundary condition $\chi(\Omega,T)=O\psi(\Omega,T)$~\cite{Werschnik2007}, as the interval $(T,T_\text{end})$ is meaningless.

\subsubsection{Method implementation and numerical convergence of Krotov algorithm}
The discontinuities given by Eq.~\eqref{eq:tk_discontinuity} can be easily implemented in the algorithm to solve the QOC equations using Krotov's algorithm~\cite{Werschnik2007}. As described in Ref.~\cite{Castro2025}, the propagation end time can be set to be the largest measurement time,~$T_\text{end}=\max(T_k)$. Hence, an observable $O(t)=O\delta(t-T_\text{end})$ sets the initial condition $\chi(\Omega,T_\text{end})=O\psi(\Omega,T_\text{end})$.

For an intermediate measurement time, $T_k\in(0,T_\text{end})$, the discontinuities only affect the propagation of the costate, see Eq.~\eqref{eq:tdse_chi}. In Krotov's method, the TDSE for the costate is solved backwards between the discontinuities or the end points. At the discontinuity $T_k$, the costate is updated to 
\begin{equation}
\label{eq:chi_update}
\chi(\Omega,T_k)\leftarrow \chi(\Omega,T_k)+O\psi(\Omega,T_k)
\end{equation}
Finally, let us remark that the convergence of Krotov's algorithm is guaranteed for $O(t)$ if it is positive definite. To prove it, we first realize that Krotov's algorithm converges for any measurement given as a time-dependent observable that it is positive definite~\cite{Ohtsuki2004}. Now, let us take
\begin{equation}
    \label{eq:obs_k_windows}
O_\tau(t)=\sum\limits_{k=0}^{N_\text{obs}}O_kf_k(t-T_k,\tau), 
\end{equation}
so that $O(t)=\lim\limits_{\tau\rightarrow 0}O_\tau(t)$ is positive definite. Hence, Krotov algorithm converges for $O_\tau(t)$ for all $\tau$, including $\tau\rightarrow 0$, which is the case under study.

\subsection{Heisenberg XXX model for a spin chain}
\label{sec:heisenbergxxx}
We apply the intermediate time control to a 5 spin-$\frac{1}{2}$ chain modeled by the Heisenberg XXX model restricted to single excitation,~\ie, 4 spins up and 1 spin down.  The control field is applied on the first particle of the chain along the interparticle axis, set as the $z$-axis for convenience. In the present case, it reads as
\begin{equation}
\label{eqn:hamiltonian}
H(t)=-\sum_{j=1}^4\vec{S}_j\cdot\vec{S}_{j+1}-\varepsilon(t)S_{z,1}
\end{equation}
 Since $  [H, S_z] = 0  $, the Hilbert space can be restricted to the Dicke subspace of a single spin excitation, spanned by the states $\ket{k}$ ($k = 1, 2, 3, 4, 5$), where $\ket{k}$ denotes the Dicke state with $4$ spins up except $k$th spin which is down~\cite{Castro2025}.
The eigenenergies of the field-free 5-spin chain for the Dicke states manifold are
collected in Table~\ref{tab:eigenenergies_eigenfunctions}.
    \begin{table*}
    \caption{\label{tab:eigenenergies_eigenfunctions} Eigenfunctions and eigenenergies of the Heisenberg XXX 5-spin chain in abscence of the magnetic field.}
\begin{ruledtabular}
\begin{tabular}{cc}
Eigenergies & Eigenfunctions\\
\hline
$-1$ & $\psi_1=\dfrac{1}{\sqrt{5}}\left(\ket{1}+\ket{2}+\ket{3}+\ket{4}+\ket{5}\right)$\\
$\dfrac{1}{4}\left(-1-\sqrt{5}\right)$ & $\psi_2=\dfrac{1}{\sqrt{5-\sqrt{5}}}\left[\ket{1}+\dfrac{1}{2}\left(\sqrt{5}-1\right)\ket{2}+\dfrac{1}{2}\left(1-\sqrt{5}\right)\ket{4}-\ket{5}\right]$\\
$\dfrac{1}{4}\left(1-\sqrt{5}\right)$ & $\psi_3=\dfrac{1}{\sqrt{5(3-\sqrt{5})}}\left[\ket{1}+\dfrac{1}{2} \left(-3+\sqrt{5}\right)\ket{2}+\left(1-\sqrt{5}\right)\ket{3}+\dfrac{1}{2} \left(-3+\sqrt{5}\right)\ket{4}+\ket{5}\right]$\\
$\dfrac{1}{4}\left(-1+\sqrt{5}\right)$ & $\psi_4=\dfrac{1}{\sqrt{5-\sqrt{5}}}\left[\dfrac{1}{2}\left(\sqrt{5}-1\right)\ket{1}-\ket{2}+\ket{4}-\dfrac{1}{2}\left(\sqrt{5}-1\right)\ket{5}\right]$\\
$\dfrac{1}{4}\left(1+\sqrt{5}\right)$ & $\psi_5=\dfrac{1}{\sqrt{5(3+\sqrt{5})}}\left[\ket{1}-\dfrac{1}{2} \left(3+\sqrt{5}\right)\ket{2}+\left(1+\sqrt{5}\right)\ket{3}-\dfrac{1}{2} \left(3+\sqrt{5}\right)\ket{4}+\ket{5}\right]$\\
\end{tabular}
\end{ruledtabular}
\end{table*}

In this work we aim to maximize the probability of finding the 5th particle with spin down (the rest with spin up), starting from the ground state of the subspace of one-spin excitations, $\psi_1$. This is equivalent to maximize the population of the Dicke state $\ket{5}$. 
In the eigenstate basis, $\ket{5}$ reads as
\begin{widetext}
\begin{eqnarray}
\nonumber   
\ket{5}&=&\dfrac{1}{\sqrt{5}}\psi_1-\dfrac{1}{\sqrt{5-\sqrt{5}}}\psi_2+\dfrac{1}{\sqrt{5(3-\sqrt{5})}}\psi_3-\dfrac{1}{2}\sqrt{\dfrac{5-\sqrt{5}}{5}}\psi_4+\dfrac{1}{\sqrt{5(3+\sqrt{5})}}\psi_5\\
    \label{eq:5th_state}
    &\approx& 0.195\psi_1-0.512\psi_2+0.632\psi_3-0.512\psi_4+0.195\psi_5 \ .
\end{eqnarray}
\end{widetext}

As the $\ket{5}$ state is not an eigenstate, its preparation at different times implies a competition between excitation and dephasing, changing the nature of revivals.

\section{Results}
\label{sec:results}
In this section, we use QOC to maximize the population of the Dicke state $\ket{5}$, $C_5(t)=|\langle \psi(t)|\ket{5}|^2$, at two measurement times and compare it with the maximization at the final time. First we set the envelope for the optimal field, $S(t)$ in Eq.~\eqref{eq:field_ele}, as
\begin{equation}
\label{eqn:envelope_field}
S(t)=
    \begin{cases}
     \sin^2\left(\dfrac{\pi t}{2t_\text{on}}\right),\quad 0\le t\le t_\text{on}  \\
     \\
     1,\quad t_\text{on}\le t\le T_\text{end}-t_\text{off}\\
     \\
     \sin^2\left(\dfrac{\pi (T_\text{end}-t)}{2t_\text{off}}\right),\quad T_\text{end}-t_\text{off}\le t\le T_\text{end}  \\
    \end{cases}
\end{equation}
where $t_\text{on}=10$ and $t_\text{off}=5$ for all the cases under study whereas $T_\text{end}$ is specified for each case. The penalty factor is $\alpha=10^{-1}$ so that the interaction with the field can be understood in terms of transitions among field-free states.

\subsection{Convergence to instantaneous measurement}
\label{sec:convergence_instantaneous_measurement}
First, we check that instantaneous measurement corresponds to a vanishing width measurement window. To do so, we compare the results obtained with the ideal instantaneous measurement ($\tau\rightarrow 0$) 
with those obtained with window functions
 against those with window functions characterized by a finite width $\tau$. Specifically, we maximize the population of $\ket{5}$ at $T_{\rm int} = 100$ and $T_{\rm end} = 200$ for FWHM $\tau=0.5,\, 2,\,5,\,10$ at each measurement time. We use the window function of finite width $\tau$,
\begin{equation}
    \label{eq:finite_window_function}
    F(t,T_{\rm int},T_{\rm end},\tau)= f(t - T_{\rm int}, \tau)+ f(t - T_{\rm end}, \tau)
\end{equation}
with
\begin{equation}
    \label{eq:sin2_window}
    f(t - T, \tau) = 
    \begin{cases}
    0,\hspace{\fill} t\le T-\tau\\
        \dfrac{1}{\tau}\cos^2\left( \dfrac{\pi (t - T)}{2\tau} \right),\qquad \hspace{\fill} T-\tau<t< T+\tau\\
        0,\hspace{\fill}  t\ge T+\tau
    \end{cases}
\end{equation}
Specifically, we maximize the population of $\ket{5}$ at $T_{\rm int} = 100$ and $T_{\rm end} = 200$ for FWHM $\tau=0.5,\, 2,\,5,\,10$ at each measurement time.
Let us note that $f(t-T,\tau)$ converges to $\delta(t-T)$ for $\tau\rightarrow 0$, so that the instantaneous measurement case corresponds to short finite width window, as already proven for the single measurement time case~\cite{Castro2025}. 

\begin{figure}
\includegraphics[width=.95\linewidth]{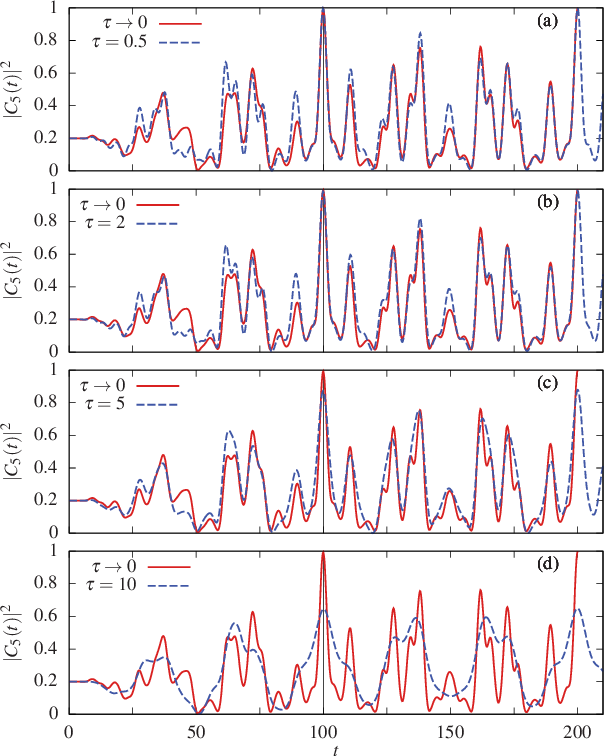}
\caption{Population of $\ket{5}$ of a 5-spin chain as a function of time for an intermediate measurement at $T_{\rm int}=100$ and final propagation time $T_{\rm end}=200$. Results are shown for measurement windows with (a) $\tau=0.5$, (b) $2$, (c) $5$ and (d) $10$ compared with the limiting case $\tau\to 0$. The penalty factor and field envelope are given by Eq.~\eqref{eqn:envelope_field}.}
\label{fig:fig1}
\end{figure}

\begin{figure}[h]
\includegraphics[width=.95\linewidth]{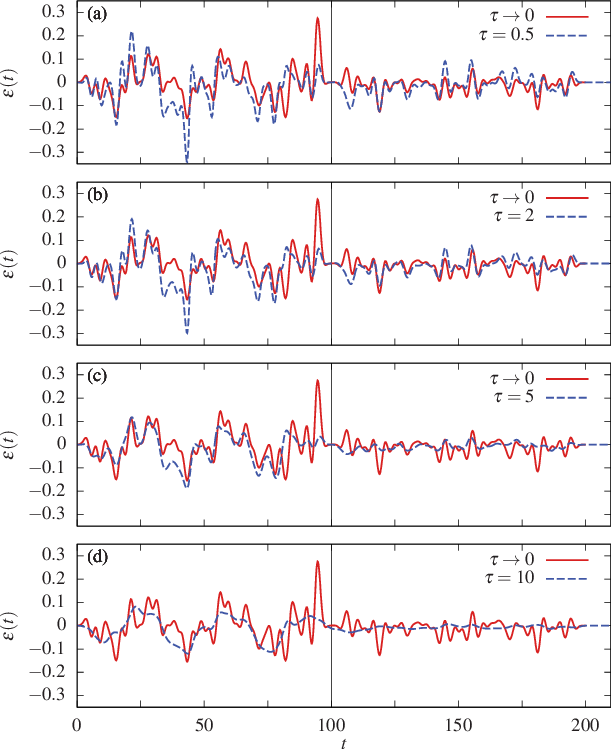}
\caption{\label{fig:fig2} Optimal control field as a function of time for an intermediate measurement at $T_{\rm int}=100$ and final propagation time $T_{\rm end}=200$. Results are shown for measurement windows with (a) $\tau=0.5$, (b) $2$, (c) $5$ and (d) $10$ compared with the limiting case $\tau\to 0$. The penalty factor and field envelope are given by Eq.~\eqref{eqn:envelope_field}.}
\end{figure}

We illustrate the convergence in the limit $\tau \to 0$ of both the population dynamics and the optimal control fields by comparing the finite-width window results with the Dirac-delta case in Figs.~\ref{fig:fig1} and~\ref{fig:fig2}, respectively. For all considered values of $\tau$, the population of state $\ket{5}$ exhibits clear peaks precisely at the maxima of the window functions at the intermediate time $T_{\rm int}$ and at the final time $T_{\rm end}$. As expected, the maximum value of the expectation value decreases with increasing $\tau$. This behavior arises because the finite-width window averages the population over a temporal interval; achieving a high averaged value requires the instantaneous population to remain elevated over a broader region away from the window peak, which constrains the maximum attainable value at the center. A similar behavior was reported for single measurements in Ref.~\cite{Castro2025}.
As the window width $\tau$ is reduced, the expectation value at the window maxima increases and converges smoothly toward the Dirac-delta limit. At $T_{\rm end}$, for instance, the population of $\ket{5}$ reaches approximately $0.999$, $0.878$, and $0.649$ for $\tau \to 0$, $\tau=5$, and $\tau=10$, respectively; at $T_{\rm int}=100$, the optimal value for $|C_5|^2$ are $0.993$, $0.870$ and $0.644$. The positions and relative heights of the main peaks, including secondary maxima, nicely converge with decreasing $\tau$. However, the finite-width results do not converge uniformly in all times. For example, the secondary peak near $t\approx 75$ still differs noticeably for $\tau=10$, while the overall shape becomes progressively closer for $\tau=5$ and thinner windows. Even for the smallest widths examined ($\tau=2$ and $\tau=0.5$), the population dynamics converge closely but do not coincide exactly with the $\tau\to 0$ case. We observed the same discrepancy in single time measurement treatment~\cite{Castro2025}, which may be related to numerical instabilities associated with the non-continuous character of the Dirac-delta case. 

Note that while reducing the window width $\tau$ yields higher peak values, a larger width can result in a greater expectation value at times off the local maxima. 
This fact highlights the practical role of measurement timing uncertainty in experimental control: when temporal resolution is limited, broadening the measurement window may offer greater robustness against timing fluctuations.
Conversely, if a target observable is not asymptotic during the process, the maximum observed value can deviate based on ideal, instantaneous measurements ($\tau \to 0$).

Now, we examine the optimal driving fields for several widths and observe that the field maximum intensity diminishes significantly following the intermediate measurement time, $T_{\text{int}}$. This reduction reflects the shifting objectives of the control protocol. In the initial stage, the field must drive the system from the ground state $\psi_1$ to $|5\rangle$, necessitating a simultaneous redistribution of eigenstate populations and tuning of quantum phases. After $T_{\text{int}}$, the populations are already near their target values; consequently, the role of the field at later times is to compensate the dephasing caused by energy level differences, see Table~\ref{tab:eigenenergies_eigenfunctions}.
\begin{figure}
    \centering
    \includegraphics[width=\linewidth]{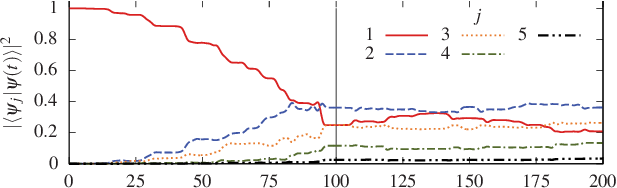} 
    \caption{\label{fig:fig3} Population of the eigenstates $\psi_k (k=1,\ldots 5)$ as a function of time, for an intermediate measurement at $T_{\rm int}=100$ and final propagation time $T_{\rm end}=200$.}
\end{figure}

In Fig.~\ref{fig:fig3}, we follow the dynamics via population histories for the optimization of the population of $\ket{5}$ at $T_{\rm int} = 100$ and $T_{\rm end} = 200$.
First, the population of the ground state, $\psi_1$, decreases stepwise to populate $\psi_2$ and $\psi_3$. Population in $\psi_4$ begins at $t \sim 75$, while it remains very low in $\psi_5$ for all times. From $t=95$ onwards, the field is so weak that the populations remain almost constant. At $T_{\text{int}}=100$, $|C_5|^2\approx 0.993$ and the populations of the eigenstates are $0.249$, $0.361$, $0.249$, $0.116$ and $0.245\cdot 10^{-1}$ in agreement with the square of the coefficients of the target state $\ket{5}$ in Eq.~\eqref{eq:5th_state}. As previously described, although the eigenstate populations are constant in the absence of the field, the population of the target state $\ket{5}$ varies. Hence, it will not be a maximum at $T=200$ unless a driving field is applied. As observed in Fig.~\ref{fig:fig2}, the field after $T_\text{int}=100$ is much weaker, ensuring that the populations do not vary, whereas the relative phases among the states are such as to synthesize state $\ket{5}$. At $T_\text{end}$, the population of the eigenstates are  $0.208$, $0.362$, $0.264$, $0.134$ and $0.326\cdot 10^{-1}$, corresponding to $|C_5|^2=0.9989$.
\begin{figure}
    \centering
    \includegraphics[width=\linewidth]{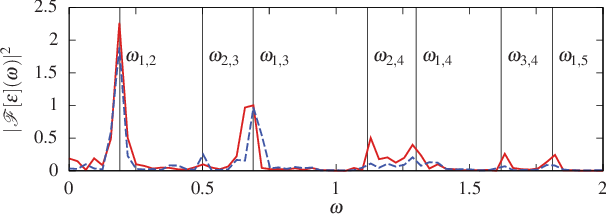}
    \caption{\label{fig:fig4} Power spectrum of the optimal driving field, $\varepsilon(t)$, for an intermediate measurement at $T_{\rm int}=100$ and final propagation time $T_{\rm end}=200$ (solid red) and single measurement at $T_{\rm end}=200$ (dashed blue). The transition energies, $\omega_{j,k}$ corresponding to the gap between the eigenenergies (see Table~\ref{tab:eigenenergies_eigenfunctions}) are shown as vertical lines.}
\end{figure}

We now discuss briefly how the power spectrum changes when incorporating an 
intermediate measurement time, as displayed in Fig.~\ref{fig:fig4}. 
We compare the spectrum obtained by maximizing the population of the target state at both $T_\text{int}=100$ and $T_\text{end}=200$ with the one obtained maximizing the population at $T_\text{end}=200$ only. 
In both cases, the spectral peaks coincide with the transition energies, particularly those involving the ground state. This fact is expected, since all transitions are allowed by the coupling term $S_{z,1}$. 

However, transitions from the second and third excited states, specifically 
$\omega_{2,3}$, $\omega_{2,4}$, and $\omega_{3,4}$, are notably more intense 
in the optimization of the two times scheme. To explain this fact, we consider the control before and after $T_\text{int}$: Initially, because the entire population is in the ground state, 
the algorithm find pulses that drive transitions directly from $\psi_1$. At the intermediate time 
$T_\text{int}$, the quantum phases must be readjusted, but now the wave function is a superposition state. Rather than relying purely on 
ground state pathways, after $T_\text{int}$ the optimization algorithm exploits all possible transitions from the wavepacket itself. Consequently, transitions originating from the second 
and third eigenstates become more valuable to the control mechanism, leading to 
the observed increase in their peak intensities relative to the values observed in the optimal pulse from the single time 
optimization scheme.

\subsection{Intermediate Measurement Times: Postpulse Field free control}
\label{sec:intermediate_measurement_times}
In this section we apply the multistage control to manipulate the revivals in the post pulse dynamics. We study two cases: enhancing existing revivals at intermediate times and inducing them. To do so, we compare the optimization of the population of $\ket{5}$ at the final measurement time with the optimization at an intermediate and final measurement time. For these tests we take $T_{\rm end}=200$ as final time, and the driving pulse is restricted to the time interval $t\in [0,100]$.

To evaluate the efficiency of our methodology, we will apply the QOC at intermediate times corresponding to revivals observed when applying QOC only at a final time. 
\begin{figure}
\includegraphics[width=.95\linewidth]{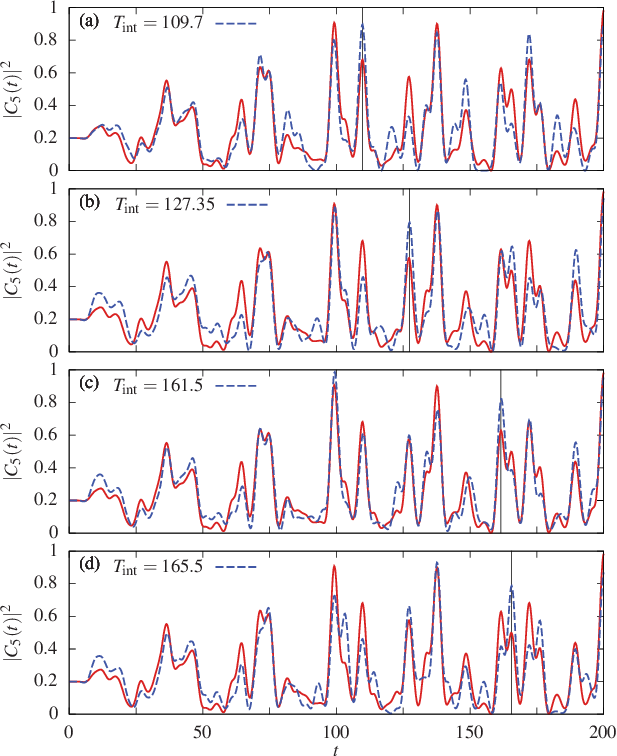}
\caption{\label{fig:fig5} Population of $\ket{5}$ as a function of time for (a) $T_\text{int}=109.7$, (b) $127.35$, (c) $161.5$ and (d) $165.5$ and $T=200$ along with non-intermediate time measurement (solid red). The penalty factor and envelope of the control field is given by Eq.~\eqref{eqn:envelope_field} and $T_\varepsilon=100$.}
\end{figure}
We first apply QOC to maximize the population of the Dicke state $\ket{5}$ 
at the final time $T_{\rm end}=200$, restricting the driving field to the interval $t \in [0,100]$. 
The resulting population dynamics of $\ket{5}$ are shown in 
Fig.~\ref{fig:fig5} (solid red), where we identify several 
maxima. We then apply the intermediate-time QOC to maximize the revivals for selected times, specifically those at $t=109.7, 127.35, 161.5,$ and $165.5$, and
the expectation value at the final time $T_{\rm end}=200$. Our results show that maximizing 
the observable at an intermediate time $T_{\text{int}}$ reduces the target state 
population at the final time $T_{\rm end}$ across all cases relative to the optimization without 
intermediate measurements. Enforcing the simultaneous maximization at two times introduces a balancing effect between the two measurements, diminishing the final maximum to enhance the average value 
with the intermediate time. 
\begin{figure}
\includegraphics[width=.95\linewidth]{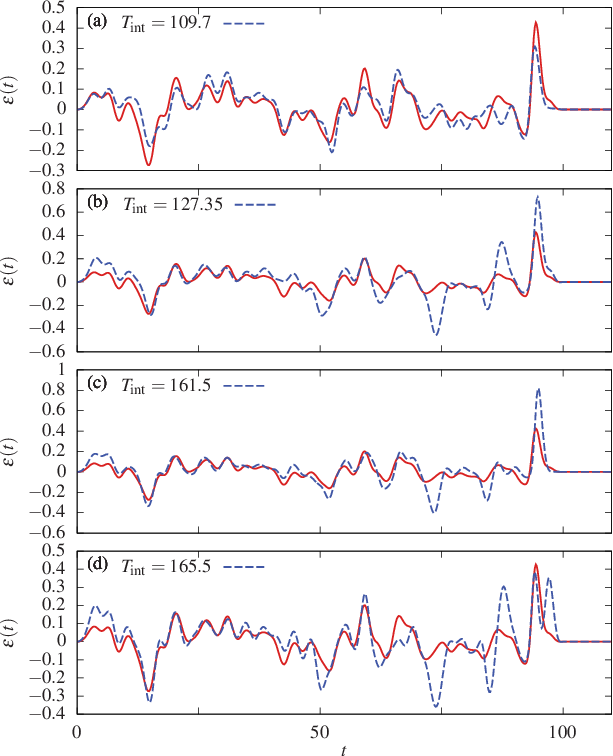}
\caption{\label{fig:fig6} Optimal field for the dynamics shown in Fig.~\ref{fig:fig5}.}
\end{figure}

We now analyze the differences among the driving fields computed via each method, as 
displayed in Fig.~\ref{fig:fig6}. The optimal field for the single time measurement scheme is characterized by several oscillations and two remarkable peaks: a local 
minimum near $t=14.6$ and a peak at $t=94.4$. We 
expect that incorporating intermediate time optimization will not drastically 
alter the pulse shape obtained from standard QOC; this is because tuning the wavepacket 
should mostly induce shifts in the relative phases while preserving the populations 
in each eigenstate, as discussed in Sec.~\ref{sec:convergence_instantaneous_measurement}. 
Indeed, this behavior holds for $t \in [0,50]$, but more pronounced discrepancies emerge 
later in the pulse. For instance, for $T_\text{int}=109.7$, the structure of the driving 
field $\varepsilon(t)$ remains almost unchanged, with the exception of the local maximum 
around $t=94.4$, which clearly splits into two. However, as the optimization 
is enforced at times further beyond the duration of driving field, these differences become more 
significant. For $T_\text{int}=127.35$, a new minimum appears at $t \approx 75$, while 
the peak at $t \approx 82$ and the final peak are heavily enhanced—the latter nearly 
doubling in amplitude. A similar trend is observed for $T_\text{int}=161.5$. Finally, 
for $T_\text{int}=165.5$, we find a more drastic modification, including the appearance 
of a new peak following the final peak of the standard QOC protocol.

\begin{figure}
\includegraphics[width=.95\linewidth]{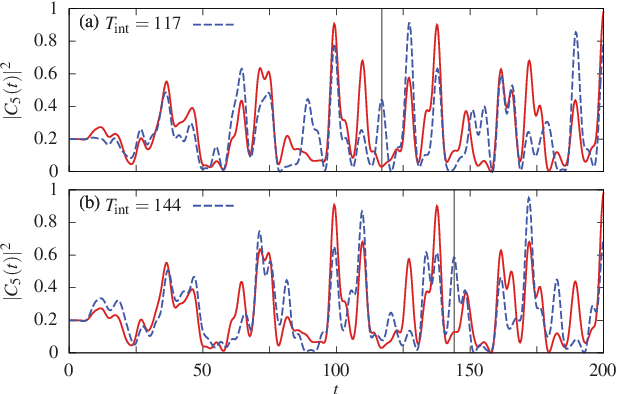}
\caption{\label{fig:fig7} Population of $\ket{5}$ as a function of time for (a) $T_\text{int}=117$ and (b) $144$ and $T=200$ along with non-intermediate time measurement (solid red). The penalty factor and envelope of the control field is given by Eq.~\eqref{eqn:envelope_field} and $T_\varepsilon=100$.}
\end{figure}

Finally, we apply the intermediate-time QOC protocol to induce post-pulse revivals at arbitrary times after the pulse has ended. Specifically, we select intermediate times where the baseline expectation value is low in the single time measurement scheme. In Fig.~\ref{fig:fig7}, we display the population dynamics of $\ket{5}$ optimized solely at $T_{\rm end}=200$, alongside the intermediate-time QOC results for $T_\text{int}=117$ and $144$. For these two intermediate times, the standard QOC values of $|C_5(t)|^2$ are approximately $0.03$ and $0.13$, respectively. In both cases, we find that the target operator is maximized at the intermediate time leading to a local maximum, reaching $0.441$ and $0.585$. However, the price to pay is high, as the maximum at the final time $T$ is greatly reduced  from around $0.982$ to $0.786$ and $0.687$, respectively. Again, the underlying reason derives from the need of the relative phases between the eigenstates to form a specific wavepacket (the target state $\ket{5}$) at different times, that are not congruent with the energy spacings. This is not possible in the weak-field limit. 

\section{Conclusions and outlook}
\label{sec:conclusions_and_outlook}

In this work, we have extended the canonical framework of Quantum Optimal Control (QOC) theory to address the optimization of multiple observables at arbitrary intermediate measurement times treated as Dirac-delta measurement windows. Using a variational approach~\cite{Castro2025}, we derived the rigorous jump conditions that govern the costate discontinuity at each measurement point. Then, using a Heisenberg XXX 5-spin chain as a testing model, we proved that this methodology can be integrated into a Krotov optimization scheme. The numerical convergence of this Dirac-delta formulation is successfully validated by proving that it represents the smooth limit of sequentially narrowing finite-width time windows.

Finally, we use our methodology to control the post-pulse, field-free population dynamics of the Dicke state $\ket{5}$ at intermediate times so that we control natural post-pulse revivals. We find that simultaneous maximization reshapes the driving field to optimize the sum of both, leading to a balance between the maximum at the final time and the intermediate target. A more pronounced effect occurs when forcing revivals at times when the standard QOC baseline population is low. In these cases, the algorithm successfully generates new local maxima, but significantly diminishes the final population. The relative phases among the eigenstates when the driving field is over determine the position of the revivals in the postpulse region. Since the algorithm is forcing the position of the revivals and the energy gaps are not commensurable, the optimal solution provides low revivals.

This multi-time control framework paves the way for advanced quantum state engineering. The capability to manipulate the system state at intermediate intervals, without sacrificing final target control, is highly relevant for multi-step quantum information processing and preserving coherence against environmental noise. Furthermore, the flexibility to optimize distinct observables throughout the evolution allows one to track the process dynamically, enabling the systematic bypassing of intermediate dark states to efficiently populate the target objective.

\begin{acknowledgments}
This research was supported by Grants No. PID2019-105458RB-I00, PID2021-122796NB-I00, PID2021-122839NB-I00 and PID2022-138288NB-C33, funded by the MICIU/AEI/10.13039/501100011033 and the ERDF/EU. The MATRIX-CM project is also aknowledged.
\end{acknowledgments}

\end{document}